\documentclass[a4paper]{jpconf}
\usepackage{graphicx}

\newcommand{\apj}{ApJ}
\newcommand{\apjl}{ApJL}

\newcommand{\mnras}{MNRAS}
\newcommand{\nat}{Nature}

\newcommand{\pasa}{PASA}
\newcommand{\araa}{ARA\&A}

\newcommand{\msun}{M$_\odot$}
\newcommand{\frb}{FRB~121102}

\begin{document}

\title{Fast Radio Bursts and their Possible Neutron Star Origins}

\author{J.\,W.\,T.\ Hessels$^{1,2}$}

\address{\tiny{
$^1$ ASTRON, the Netherlands Institute for Radio Astronomy, Postbus 2, 7990 AA, Dwingeloo, The Netherlands

$^2$ Anton Pannekoek Institute for Astronomy, University of Amsterdam, Science Park 904, 1098 XH Amsterdam, The Netherlands.}}

\ead{j.w.t.hessels@uva.nl}

\begin{abstract}
The discovery of the `Lorimer Burst', a little over a decade ago, ignited renewed interest in searching for short-duration radio transients\cite{Lor07}.  This event is now considered to be the first established Fast Radio Burst (FRB), which is a class of millisecond-duration radio transients\cite{Tho13}.  The large dispersive delays observed in FRBs distinguish them from the individual bright pulses from Galactic pulsars, and suggests that they originate deep in extragalactic space.  Amazingly, FRBs are not rare: the implied event rate ranges up to many thousands of events per sky, per day\cite{Cha16}.  The fact that only two dozen FRBs have been discovered to date is a consequence of the limited sensitivity and field of view of current radio telescopes\cite{Pet16}.  The precise localization of \frb, the first and currently only FRB observed to repeat\cite{Spi14,Spi16,Sch16}, has led to the unambiguous identification of its host galaxy and thus proven its extragalactic origin and large energy scale\cite{Cha17,Ten17,Mar17}.  It remains unclear, however, whether all FRBs are capable of repeating (many appear far less active \cite{Pet15a}) or whether \frb\ implies that there are multiple sub-classes.  Regardless, the repetitive nature of \frb\ and its localization to within a star-forming region in the host galaxy\cite{Bas17} imply that the bursts might originate from an exceptionally powerful neutron star -- one necessarily quite unlike any we have observed in the Milky Way.  In these proceedings, I give a very brief introduction to the FRB phenomenon and focus primarily on the insights that \frb\ has provided thus far.

\end{abstract}

\section{Introduction}
\vspace{5mm}
Radio pulsar surveys have made great strides in the last decade.  Thanks to increased computational power, higher time and frequency resolution can be achieved.  This is enabling, e.g., the discovery of many more millisecond pulsars.  Furthermore, long dwell times and larger fields-of-view are being employed to search for sporadically emitting sources like the rotating radio transients\cite{McL06}.  Larger fields of view and more on-sky time also open the prospect of discovering other types of millisecond-duration radio flashes.  For example, given the scientific importance of gamma-ray bursts\cite{Geh12}, it is natural to ponder and hope: are there also similar sirens of extreme (astro)physics to be found at radio wavelengths?

Unfortunately, there is no sufficiently sensitive all-sky radio monitor yet available.  Compared to the all-sky monitoring routinely done in X-rays and $\gamma$-rays, a typical `wide-field' radio telescope capable of offering (adequately sensitive) high-time-resolution burst searches has a field of view of only $\sim 1-10$\,sq\,deg.  In fact, the 64-m Parkes  telescope, which has found the vast majority of the roughly two dozen FRBs published to date, requires $\sim 10$\,days of observations to discover a single FRB, despite their inferred large event rate.  Again thanks to increasing computational power, however, truly wide-field radio telescopes like CHIME\cite{Ami17}, UTMOST\cite{Cal16}, ASKAP\cite{Ban17} and APERTIF\cite{Maa17} are beginning to collect high-time-resolution data, and promise to detect up to dozens of FRBs per day.  Such searches may also identify even rarer types of short-duration radio transients.

Adding to the computational challenges, short-duration radio waves are strongly affected by propagation effects induced by the intervening magneto-ionized medium between source and observer\cite{Ric90}.  In particular, dispersion causes astrophysical radio pulses to arrive later at lower frequencies.  This requires correction (`de-dispersion') in wide-band radio observations.  Such effects are not purely a hindrance, however; they can also encode valuable information about the source's properties, its local environment, and its distance\cite{Cor17}.  For example, the measurement of a relatively high Faraday rotation measure in one FRB implied that it is associated with a dense, magnetized plasma\cite{Mas15}.  Furthermore, the dispersion measure (DM$ = \int_0^d n_e(l) dl$) is the integrated column density of ionized material along the line of sight, and can be used as a proxy for distance\cite{Cor02}.  This is how the extragalactic distances of FRBs are inferred in the (majority of) cases where no direct distance measurement is available.  Dispersion is also critical for separating genuine astrophysical FRBs from a strong background of man-made radio frequency interference (RFI), particularly because some artificial signals can appear (at first glance) similar to FRBs\cite{Pet15b}.

A decade after the discovery of the  `Lorimer Burst'\cite{Lor07}, the FRB phenomenon remains enigmatic.  The field has gained great momentum, however, in the last few years, thanks in large part to many new discoveries and observational insights (see \cite{Pet17} for a broader review than provided here).  The new wide-field radio telescope `FRB factories', as well as prospects for directly localizing bursts in real time\cite{Law15}, promise major advances in the {\it coming} years as well.

Many theories for the physical nature of the FRBs have been proposed\cite{Mac15,Kat16,Pet17} --- including both cataclysmic events involving neutron star collision (or collapse), along with non-cataclysmic scenarios involving a young and/or highly magnetized neutron star.  While some FRBs appear to be one-off events --- arguably supporting a cataclysmic origin --- the sporadically repeating \frb\ has also recently been discovered \cite{Spi16}.  It is thus currently unclear whether there is a single type of FRB, or whether we are seeing multiple source populations \cite{Sch16}.  In any case, it is very plausible that exotic manifestations of neutron stars are at least part of the puzzle.  After all, several types of millisecond-duration radio emission are already known to be created in neutron star magnetospheres: i. canonical, magnetic polar cap pulsar radio emission; ii. giant pulses, like those seen in the Crab; and iii. radio pulses from magnetars, which arguably originate from yet another physical process\cite{Lor04}.  In these proceedings I tell the story of the discovery and follow-up study of `The Repeating FRB', \frb.  The repetitive nature of this particular FRB has allowed deep, multi-wavelength campaigns that have delivered many key insights.

\begin{figure}[h]
\begin{minipage}{20pc}
\includegraphics[width=20pc]{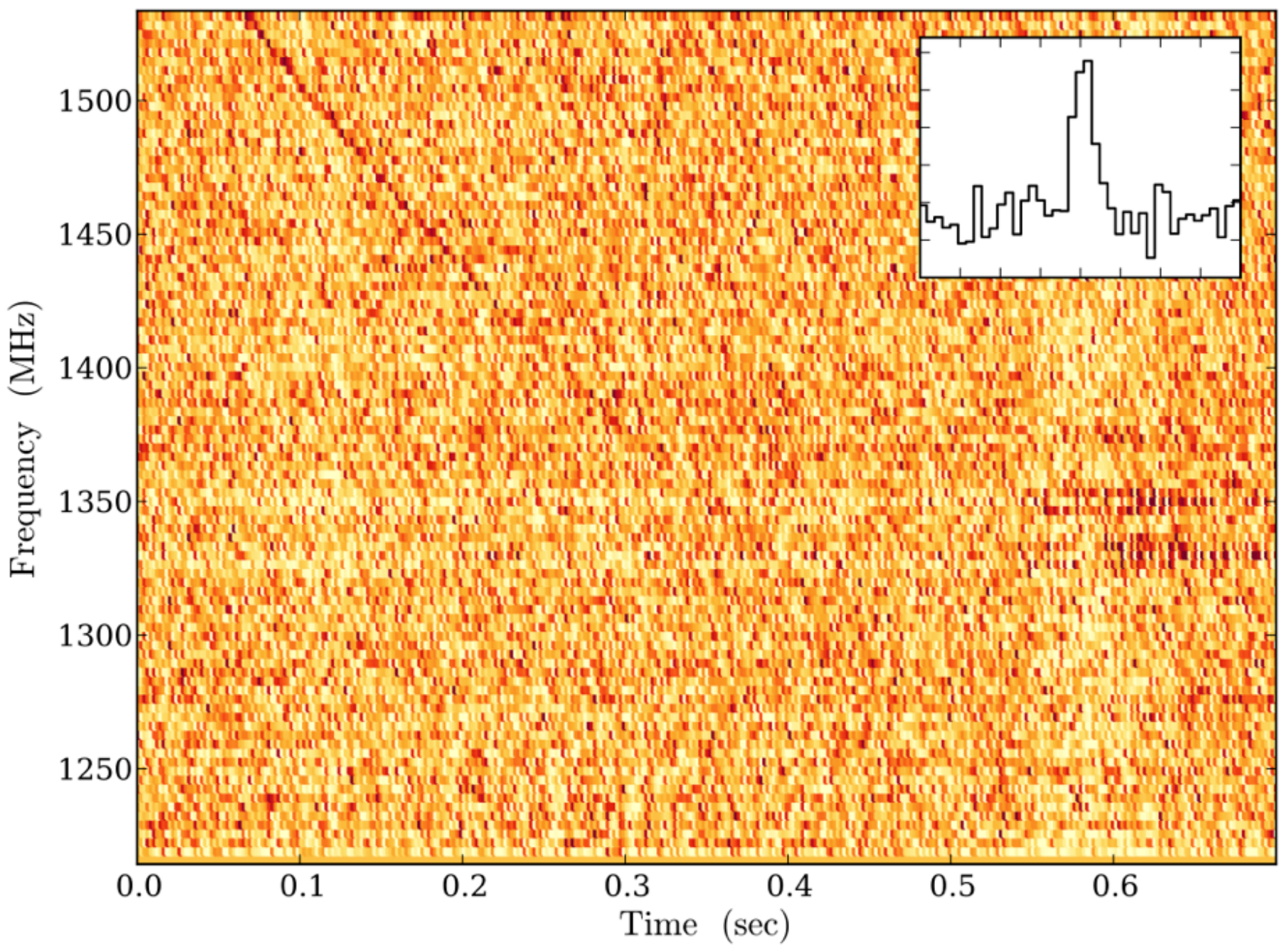}
\caption{\label{fig:spitler_burst}The first observed burst from \frb\cite{Spi14}.  The main panel shows the characteristic dispersive delay and burst intensity as a function of frequency and time.  The inset is the burst light curve after dedispersion and integrating across the observed band.  (Credit: L.~Spitler)}
\end{minipage}\hspace{2pc}%
\begin{minipage}{16pc}
\includegraphics[width=16pc]{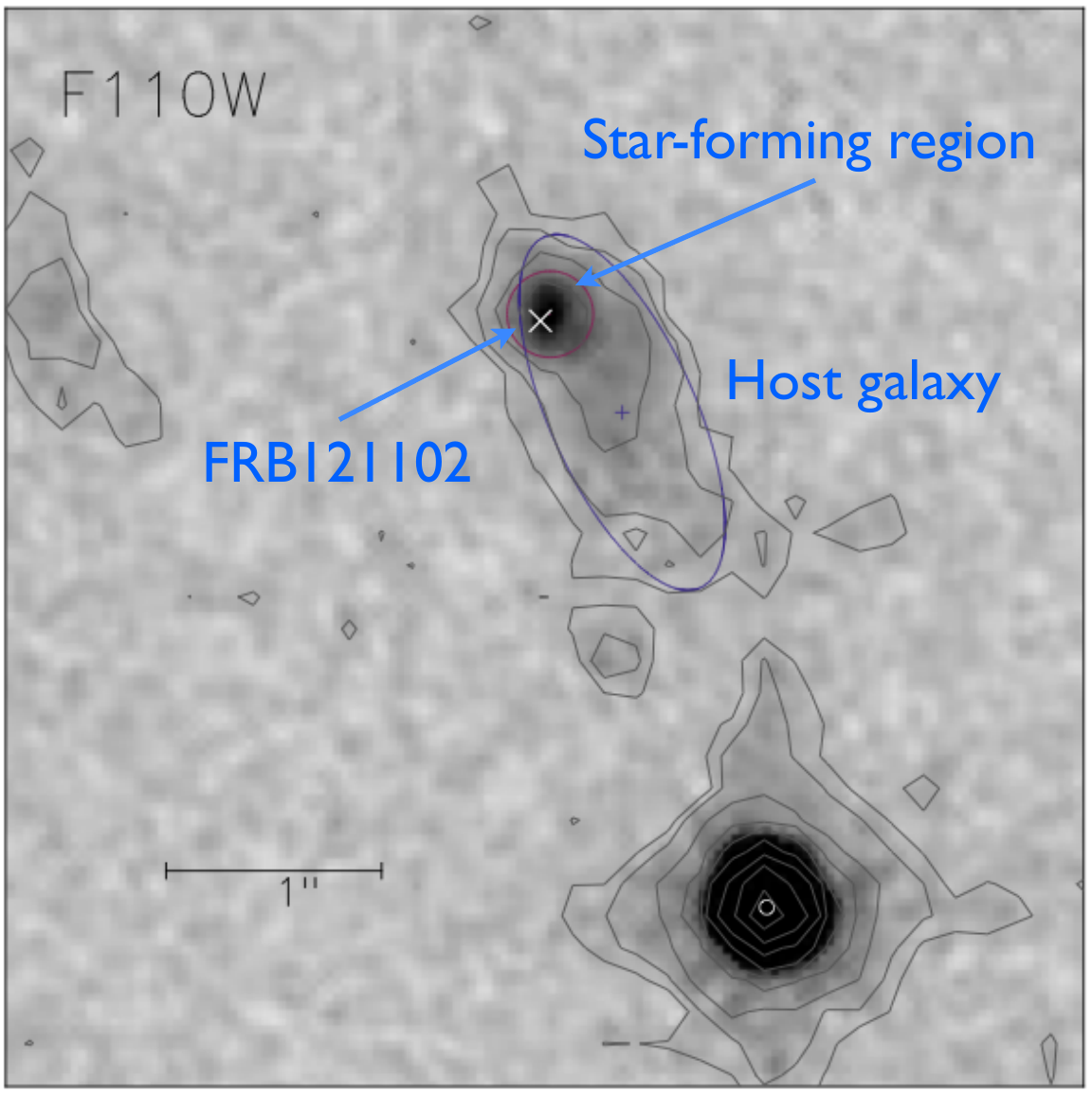}
\caption{\label{fig:frb_loc}Localization of \frb\ within a star-forming region in its host galaxy\cite{Bas17}.  The position of the EVN burst and persistent source localization is shown by the white `x'\cite{Mar17}. (Adapted from Bassa et al. 2017)}
\end{minipage} 
\end{figure}

\section{The Repeater \frb}
\vspace{5mm}
Until 2014, all reported FRBs had been discovered using the 64-m Parkes radio telescope and its multi-beam receiver.  The discovery of \frb\ (also known as the `Spitler Burst'; Figure~\ref{fig:spitler_burst}) in the PALFA pulsar survey\cite{Spi14}, which uses the 305-m Arecibo telescope\cite{Cor06,Laz15}, was the first non-Parkes detection.  At the time, the discovery greatly bolstered the astrophysical interpretation of the FRBs because it helped dispel concerns that the phenomenon was some pernicious source of local interference, like the so-called Perytons\cite{Bur11,Pet15b}.  

Surprisingly, the burst appeared to have an inverted spectral index, becoming much brighter towards the top of the observing band.  This had not been seen in previous FRBs, and was explained as possibly coming from the detection of the burst in a side-lobe of the receiver sensitivity pattern (though this required fine-tuning \cite{Kul15}).  The low Galactic latitude of the burst ($b = -0.2^{\circ}$) left open the possibility that \frb\ could be a intermittant Galactic pulsar like the rotating radio transients (RRATs)\cite{McL06}.  However, since the PALFA survey is restricted to Galactic latitudes $|b| < 5^{\circ}$, by definition it can only find such sources along the Galactic plane.  Furthermore, \frb's DM$ = 557$\,pc\,cm$^{-3}$ is roughly $3 \times$ larger than the maximum predicted Galactic contribution along this line of sight\cite{Cor02}, meaning that an unmodelled HII region would need to be invoked to explain the apparent discrepancy.  The lack of observed pulsars with anomalously high DMs in this part of the sky provided an additional argument that \frb\ was something different.

A follow-up observing campaign was planned in order to search for repeat bursts at the same sky location and DM.  Though no other FRB had been seen to repeat, despite dedicated follow-up\cite{Pet15a}, the possibility of a RRAT origin for \frb\ gave extra motivation for continued monitoring.  Surprisingly, 10 more bursts were detected in just a few hours of additional observations\cite{Spi16} (for comparison, the discovery observation lasted only $\sim 200$\,s).  In a grid of pointing positions that covered the initial discovery position and side-lobes of the receiver, these additional bursts all had consistent sky position and DM, thereby demonstrating that they were coming from the same source, and that it was following the sidereal rate.  The detection of multiple bursts immediately ruled out a cataclysmic origin for \frb; whatever source is producing the bursts needs to survive the energetic events that produce them.  Interestingly, the new sample of bursts showed a broad range of burst widths, morphologies, and spectra\cite{Spi16}.  Though the refined localization indicated that the first burst had indeed been detected in a receiver side-lobe, the spectral volatility could not be ascribed to an instrumental effect.  Rather, it would have to be intrinsic to the emission process itself or the result of propagation in the intervening magneto-ionized material.

Multi-wavelength observations of the field found no evidence for Galactic structure capable of explaining \frb's anomalously high DM\cite{Sch16}.  In tens of hours of further monitoring, however, only 6 more bursts were detected, using the Green Bank Telescope (GBT) and Arecibo.  This suggested that the initial follow-up campaign had been rather fortunate in finding 10 bursts, and that the activity level of \frb\ is highly variable (in fact 6 of the 10 bursts originally presented were detected within a single 10-min time window\cite{Spi16}).

The $\sim 2^{\prime}$ localization precision of the Arecibo redetections was completely insufficient to identify a host galaxy.  Rather, precise and direct localization of \frb\ was achieved after tens of hours of monitoring using the Very Large Array (VLA) in a fast-dump ($5$\,ms visibilities) mode.  During a period of pronounced source activity in September 2016, these observations captured 9 bursts, which were directly imaged and localized to sub-arcsecond precision\cite{Cha17}.  Unlike indirect methods that attempt to associate an FRB with a multi-wavelength counterpart or afterglow\cite{Kea16}, this method leaves no ambiguity about the association.

With a precision localization in hand, it soon became apparent that \frb\ has both persistent radio and optical counterparts.  The VLA continuum maps identified a $\sim 200$\,$\mu$Jy persistent radio source at the exact location of the bursts.  Very-long-baseline radio interferometric (VLBI) observations using the European VLBI Network (EVN) and the Very Long Baseline Array (VLBA) also detected this source, and showed that it is compact on milli-arcsecond angular scales.  Archival Keck, and newly acquired Gemini data identified a weak ($\sim 25$\,mag) optical counterpart, possibly just barely resolved.

Additional EVN observations, including Arecibo, detected one bright and three weaker bursts.  These observations stored the raw voltage signals from each radio dish, which allowed offline correlation to be done at high time resolution during the specific times of the bursts.  With a longer-baseline interferometric array, it was possible to improve the precision of the burst localization by another order of magnitude, achieving a $\sim 12$\,mas burst position uncertainty (dominated by limited {\it uv} coverage)\cite{Mar17}.  These observations also detected the persistent radio counterpart and showed that it must be $< 0.7$\,pc in extent and $< 40$\,pc from the source of the bursts.  This established that there is either a physical link between the source of the bursts and the source of the persistent radio emission or, perhaps, that the persistent radio source is {\it directly} responsible for the bursts themselves.

Using the 8-m Gemini North, optical spectroscopy of the persistent optical counterpart detected strong emission lines and was able to measure a redshift: $z = 0.193$ (luminosity distance $\sim 1$\,Gpc).  This confirmed \frb's extragalactic origin beyond any doubt, and established the first robust association of an FRB with a host galaxy\cite{Ten17}.  Importantly, this also establishes the energy scales of the bursts and persistent source: $L_{\rm Burst} \sim 10^{40-42} \times (\Delta{\Omega_b}/4\pi)$\,erg\,s$^{-1}$ and $L_{\rm Persistent} \sim 7 \times 10^{38} \times (\Delta{\Omega_p}/4\pi)$\,erg\,s$^{-1}$.  Here $\Delta{\Omega_b}$ is the burst emission solid angle (which could be small) and  $\Delta{\Omega_p}$ is the persistent emission solid angle, almost certainly different than $\Delta{\Omega_b}$, and potentially isotropic.  Surprisingly, the host galaxy is a low-metallicity dwarf with a total stellar mass $\sim 10^8$\,\msun.  Interestingly, superluminous supernovae (SLSNe) and long gamma-ray bursts (LGRBs) are also preferentially found in dwarf galaxies similar to \frb's host\cite{Mar17,Ten17,Met17}.  This presents the tantalizing possibility that \frb\ is a young and extreme neutron star created during such an explosion.  This hypothesis is now being tested via targeted searches for millisecond-durations radio bursts towards the sites of known SLSNe and LGRBs.  More recent {\it Hubble} observations show that \frb\ is close to the center (Figure~\ref{fig:frb_loc}) of an active star-forming region (or the blend of several unresolved star-forming regions)\cite{Bas17}.  This accounts for the host galaxy's optical emission lines, and further strengthens the association of \frb\ to (massive) star formation.

While the repetition of \frb\ has thus had huge practical advantages for its localization, this also affords deep, multi-wavelength observing campaigns that can more precisely characterize the properties of the radio bursts themselves and search for prompt emission at higher energies.  Using coherent dedispersion, it is possible to recover a much higher effective time resolution, which does not suffer from dispersion smearing within individual frequency channels.  This has demonstrated that \frb\ bursts are often multi-peaked and can have structure at $< 1$\,ms timescales (\cite{Spi16,Sch16}; also Hessels et al., in prep.).  While the bursts are often narrow-band ($\sim 100-200$\,MHz at central observing frequencies of $\sim 1.4$\,GHz), multi-telescope observations show that they are also occasionally broadband (spanning several GHz)\cite{Law17}.   The degree to which this is intrinsic to the burst emission mechanism or due to extrinsic propagation effects remains unclear.  While there are certainly precedents for narrow-band features in the bursts of the Crab pulsar\cite{Han07} it has also been suggested that the spectral features can be due to plasma lensing, which has the added advantage that it can boost the observed brightness of the bursts and lessen the energy required to explain the bursts\cite{Cor17}.  Multi-wavelength observations, simultaneous with radio burst detections, have thus far failed to detect optical or X-ray counterparts to the bursts or persistent radio source\cite{Har17,Sch17}.  Unfortunately, given the large distance to \frb, this is only mildly constraining on possible magnetar models\cite{Sch17}.

In summary, there are several lines of evidence that suggest that \frb\ originates from a young and highly magnetized neutron star.  Several models invoking `super-giant' radio pulses, magnetar flares, etc. have been proposed\cite{Pen15,Con16,Cor16,Lyu14}.  While the precedents of millisecond-duration radio pulses from neutron stars make that a logical (and conservative) conclusion, the extreme energetics and lack of direct Galactic analogue leave various open questions: e.g. are the \frb\ bursts rotationally or magnetically powered?  Models have also been proposed in which a neutron star can produce sufficiently luminous bursts only because of an external influence (e.g. \cite{Zha17}).  Ultimately, one should remain open-minded because a definitive proof of the neutron star scenario is still lacking: e.g. the detection of a well-defined periodicity that could only plausibly be ascribed to a rotating neutron star.

An equally compelling puzzle is whether \frb\ has the same physical origin as the remaining FRB population\cite{Spi16,Pet16}.  Deep searches for repeats from other FRBs have so far detected no additional bursts\cite{Pet15a}, and thus \frb\ has to (at least) be atypically active compared to other sources; this requires explanation either via intrinsic source properties or external effects.  That said, \frb\ shares many of the other characteristics of the general population, and the spectral and temporal properties of other FRBs suggest some commonality\cite{Cha16,Rav16}.  There are high chances that more repeating FRBs will be discovered in the coming years by the aforementioned `FRB factories', and that FRBs will begin to be precisely localized in real time.  Thus, it appears that clarity is within sight.

\section*{Acknowledgements}
I heartily thank my various collaborators for their great FRB research accomplishments in the last few years.  In particular, I would like to thank the following colleagues for their help in preparing the talk on which this proceedings contribution is based; in alphabetical order: Cees Bassa, Shami Chatterjee, Jim Cordes, Kelly Gourdji, Casey Law, Benito Marcote, Daniele Michilli, Zsolt Paragi, Emily Petroff, Paul Scholz, and Laura Spitler.  On a personal note, I also thank the dedicated staff of the Arecibo Observatory for their continued professionalism and hard work -- especially in the aftermath of a major natural disaster in Puerto Rico.

\section*{References}

\providecommand{\newblock}{}

\end{document}